\documentclass[11pt]{article}
\usepackage{moriond,epsfig,amsmath,amssymb,bm}

\def\nbslash{\rlap{\hspace{0.02cm}/}{\bar n}}

\begin{document}
\vspace*{4cm}
\title{DIRECT CP ASYMMETRY IN  \boldmath $\bar B\to X_{s,d}\,\gamma$ DECAYS}

\author{GIL PAZ}

\address{Enrico Fermi Institute  \\
The University of Chicago, Chicago, Illinois, 60637, USA}

\maketitle\abstracts{The CP asymmetry in inclusive $\bar B\to X_{s,d}\,\gamma$  decays is an important probe of new physics. The theoretical prediction was thought to be of a perturbative origin, and in the standard model, to be about 0.5 percent. In a recent work with M. Benzke, S.J. Lee and M. Neubert, we have shown that the asymmetry is in fact dominated by non-perturbative effects. Since these are hard to estimate, it reduces the sensitivity to new physics effects. On the other hand, these new non-perturbative effects suggest a new test of new physics by looking at the difference of the CP asymmetries in charged versus neutral B-meson decays. }

\section{Introduction}
Inclusive radiative B decay modes such as $\bar B\to X_{s}\,\gamma$  do not occur in the Standard Model (SM) at tree-level. They can only take place via  loop suppressed processes. At low energies, one can describe the decay $\bar B\to X_{s}\,\gamma$ via the operator  $Q_{7\gamma}=(-e/8\pi^2)m_b \bar s \sigma_{\mu\nu}F^{\mu\nu}(1+\gamma_5)b$, which appears in the effective Hamiltonian as ${\cal H}_{\rm eff}\ni -(G_F/\sqrt{2}) \lambda_tC_{7\gamma}Q_{7\gamma}$. The decay $\bar B\to X_{s}\,\gamma$ can also receive contributions from loops that contain particles which appear in extensions of the SM, making it an important probe of new physics effects. Such effects would only modify the Wilson coefficient  $C_{7\gamma}$ in the effective Hamiltonian. 

But $Q_{7\gamma}$ is not the only operator that contributes to $\bar B\to X_{s}\,\gamma$. Instead of producing  the photon directly, we can produce a gluon or a quark pair and convert them to a photon. In other words, the contribution of operators such as $Q_{8g}=(-g/8\pi^2)m_b {\bar s} \sigma_{\mu\nu}G^{\mu\nu}(1+\gamma_5){ b}$ and $Q_1^c=(\bar c{ b})_{V-A}({\bar s} c)_{V-A}$ is also important. Such a conversion usually ``costs us" either a factor of $\alpha_s$ from a loop that contains a ``hard" gluon, or a factor of $\Lambda_{\rm QCD}/m_b$ from a production of extra ``soft" particles in the conversion process. In summary, in order to describe inclusive radiative B decays we need the full effective Hamiltonian 
\begin{equation}\label{Heff}
{\cal H}_{\rm eff}=\frac{G_F}{\sqrt{2}}\sum_{p=u,c}\lambda_p\left(\vspace{-2em} C_1 Q_1^p+C_2 Q_2^p+\sum_{i}\vspace{-2em} C_iQ_i+C_{7\gamma}Q_{7\gamma}+C_{8g}Q_{8g}\right)+{\rm h.c.}\,,
\end{equation}
where $\lambda_p=V^*_{pb}V_{ps}$ ($\lambda_p=V^*_{pb}V_{pd}$) for  $\bar B\to X_{s}\,\gamma$  ($\bar B\to X_{d}\,\gamma$) decays. The most important operators, due to their larger Wilson coefficients, are $Q_{7\gamma},Q_{8g},$ and $Q_1$. From the effective Hamiltonian one can calculate various observables, in particular the CP asymmetry which is the topic of this talk.

The measured CP asymmetry in  $\bar B\to X_{s}\,\gamma$, as given by the Heavy Flavor Averaging Group, is \cite{TheHeavyFlavorAveragingGroup:2010qj}
\begin{equation}
 {\cal A}_{X_s\gamma} 
   = \frac{\Gamma(\bar B\to X_s\gamma)-\Gamma(B\to X_{\bar s}\gamma)}   
          {\Gamma(\bar B\to X_s\gamma)+\Gamma(B\to X_{\bar s}\gamma)} 
   = - (1.2\pm 2.8)\%\,.
 \end{equation}
This result is based on averaging of BaBar, Belle, and CLEO measurements, which actually measure ${\cal A}_{X_s\gamma}(E_\gamma\ge E_0)$ with $1.9\le E_0\le 2.1$\,GeV. 

What about the theoretical prediction?  It is well known that in order to observe CP violation we should have an interference of two amplitudes that differ both in their ``weak" (CP odd) and ``strong" (CP even) phases. The weak phases can appear from the CKM matrix elements in (\ref{Heff}), or from complex Wilson coefficients in (\ref{Heff}). The latter source cannot contribute in the SM, where the Wilson coefficients are real.  The strong phases can arise, for example, from loops, and as such are $\alpha_s$ suppressed. Indeed, the theoretical prediction for the CP asymmetry was thought to be of a perturbative origin.  The perturbative theoretical  prediction is given by \cite{Kagan:1998bh,Asatryan:2000kt}
\begin{equation}\label{direct}
\begin{aligned}
   &{\cal A}_{X_s\gamma}^{\rm dir}(E_0)
   = \alpha_s\,\Bigg\{ \frac{40}{81}\,\mbox{Im}\,\frac{C_1}{C_{7\gamma}} 
    - \frac{8z}{9}\,\Big[ v(z) + b(z,\delta) \Big]\,
    \mbox{Im}\bigg[(1+\epsilon_s)\,\frac{C_1}{C_{7\gamma}}\bigg] \\
   &\hspace{-0.5em} - \frac49\,\mbox{Im}\,\frac{C_{8g}}{C_{7\gamma}}+ \frac{8z}{27}\,b(z,\delta)\,
    \frac{\mbox{Im}[(1+\epsilon_s)\,C_1 C_{8g}^*]}{|C_{7\gamma}|^2}
    + \frac{16z}{27}\,\tilde b(z,\delta)\,\bigg|\frac{C_1}{C_{7\gamma}}\bigg|^2\,
    \mbox{Im}\,\epsilon_s \Bigg\} \,,
\end{aligned}
 \end{equation}
where  $\delta=(m_b-2E_0)/m_b$, $z=(m_c/m_b)^2$ and  $\epsilon_s=(V_{ub} V_{us}^*)/(V_{tb} V_{ts}^*)\approx \lambda^2(i\bar\eta-\bar\rho)$. One can simplify this expression by taking  $m_c^2={\cal O}(m_b\Lambda_{\rm QCD})$, and expanding  in $z,\delta={\cal O}(\Lambda_{\rm QCD}/m_b)$. In this limit \cite{Benzke:2010tq}
\begin{equation}
   {\cal A}_{X_s\gamma}^{\rm dir}
   = \alpha_s\,\Bigg\{ \frac{40}{81}\,\mbox{Im}\frac{C_1}{C_{7\gamma}}
    - \frac49\,\mbox{Im}\frac{C_{8g}}{C_{7\gamma}} 
  - \frac{40\Lambda_c}{9m_b}\,\mbox{Im}\bigg[(1+\epsilon_s)\,
    \frac{C_1}{C_{7\gamma}}\bigg] 
    + {\cal O}\bigg( \frac{\Lambda_{\rm QCD}^2}{m_b^2} \bigg) \Bigg\} \,,
    \end{equation}
where  $\Lambda_c(m_c,m_b) \approx 0.38\,\mbox{GeV}$. In the  
 SM, where $C_i$ are real,  we find a triple suppression: from $\alpha_s$, from $\mbox{Im}(\epsilon_s)\sim\lambda^2\approx0.05$, and from $(m_c/m_b)^2\sim\Lambda_{\rm QCD}/m_b$. All together, the theoretical prediction for the SM is an asymmetry of about $0.5\%$ \cite{Soares:1991te,Kagan:1998bh,Ali:1998rr}. A dedicated analysis  \cite{Hurth:2003dk}  finds 
 ${\cal A}_{X_s\gamma}^{\rm SM}=(0.44_{\,-\,0.10}^{\,+\,0.15}\pm 0.03_{\,-\,0.09}^{\,+\,0.19})\%$,
 where the errors are from  $m_c/m_b$, CKM parameters, and scale variation, respectively.  
 
Comparing the theoretical prediction to the measured value, ${\cal A}_{X_s\gamma} = - (1.2\pm 2.8)\%$, there is room for new physics effects. There are many studies of such effects on the CP asymmetry, see for example the references within \cite{Benzke:2010tq}. 
From the experimental side, reducing the experimental error below the $1\%$ level is one of the goals of the future B factories \cite{Bona:2007qt,Aushev:2010bq}. There is a problem, though, since the theoretical prediction is not complete...

\section{Resolved Photon Contributions} 
The theoretical prediction of (\ref{direct}) is missing the resolved photon contributions \cite{Lee:2006wn,Benzke:2010js}. Unlike the direct photon contribution in which the photon couples to a local operator mediating the weak decay, the resolved photon contribution arise from indirect production of the photon, accompanied by other soft particles.  The resolved photon contributions give rise to non-perturbative $ {\cal O}(\Lambda_{\rm QCD}/m_b)$ corrections to  $\Gamma(\bar B\to X_s\,\gamma)$. This is very different from other inclusive B decays, such as $\bar B\to X_{c,u} l\,\bar\nu$, where the non-perturbative corrections to the total rate are ${\cal O}( \Lambda^2_{\rm QCD}/m^2_b)$. The resolved photon contributions give  the largest error of about $5\%$ on the total $\bar B\to X_s\,\gamma$ rate.  

How important are they for the CP asymmetry? For the total rate the direct photon contributions are an ${\cal O}(1)$ effect, while the resolved photon contributions are $ {\cal O}(\Lambda_{\rm QCD}/m_b)$ suppressed. For the CP asymmetry the direct photon contribution discussed above are $\alpha_s$ suppressed, so the resolved photon contributions can give a potentially large effect. 

Before giving explicit expressions  for the resolved photon contributions to the CP asymmetry, let us comment on their structure. Schematically, the resolved photon contributions appear in the form of ${\cal A}_{X_s\gamma}^{\rm res} \sim\bar{J}\otimes h$.  The functions $\bar{J}$ can be calculated in perturbation theory. The soft functions $h$ are matrix elements of non-local operators. They cannot be extracted from data, and must be modeled. How do the strong phases arise for the resolved photon contributions? It can be shown \cite{Benzke:2010js} that the functions $h$ are real by using parity, time reversal, and heavy quark symmetry. The functions $J$, on the other hand, are complex since they arise from uncut propagators and loops.  

At the lowest order in $\alpha_s$ and $\Lambda_{\rm QCD}/m_b$, the resolved photon contribution to the CP asymmetry is
\begin{equation}\label{resolved}
 {\cal A}_{X_s\gamma}^{\rm res}
   = \frac{\pi}{m_b}\,\bigg\{
    \mbox{Im}\bigg[(1+\epsilon_s)\,\frac{C_1}{C_{7\gamma}}\bigg]\,{ \tilde\Lambda_{17}^c}
    - \mbox{Im}\bigg[\epsilon_s\,\frac{C_1}{C_{7\gamma}}\bigg]\,{\tilde\Lambda_{17}^u} + \mbox{Im}\,\frac{C_{8g}}{C_{7\gamma}}\,
    4\pi\alpha_s\,{\tilde\Lambda_{78}^{\bar B}} \bigg\} \,,
\end{equation}
with
\begin{equation}
\begin{aligned}
{ \tilde\Lambda_{17}^u }&= \frac23\,{\ h_{17}(0)}\\
{   \tilde\Lambda_{17}^c }
   &= \frac23 \int_{4m_c^2/m_b}^\infty\!\frac{d\omega_1}{\ \omega_1}\,
  {  f\bigg( \frac{m_c^2}{m_b\,\omega_1} \bigg)}\,{\ h_{17}(\omega_1)} \\
{\    \tilde\Lambda_{78}^{\bar B}}
   &= 2\int_{-\infty}^\infty\!\frac{d\omega_1}{\ \omega_1}
    \left[ {\ h_{78}^{(1)}(\omega_1,\omega_1) - h_{78}^{(1)}(\omega_1,0)} \right]\,, 
\end{aligned}
\end{equation}
where 
$ f(x) = 2x\ln[(1+\sqrt{1-4x})/(1-\sqrt{1-4x})]$.
The soft functions $h_{ij}$ are in light-cone gauge $\bar n\cdot A=0$,
\begin{eqnarray}\label{eqn:h17}
h_{17}(\omega_1,\mu) 
   &=& \int\frac{dr}{2\pi}\,e^{-i\omega_1 r}\,
    \frac{\langle\bar B| \bar h (0)\,
    \nbslash\,i\gamma_\alpha^\perp\bar n_\beta\,
    g G_s^{\alpha\beta} 
    (r\bar n)\,
     h(0) |\bar B\rangle}{2M_B}  \\
 h_{78}^{(1)}(\omega_1,\omega_2,\mu) 
   &=& \int\frac{dr}{2\pi}\,e^{-i\omega_1 r}\!
    \int\frac{du}{2\pi}\,e^{i\omega_2 u} 
    \frac{\langle\bar B| \bar h (0)\,T^A\,
          \nbslash\,
           h(0)\,
          \sum{}_q\,e_q\,\bar q (r\bar n)\,
          \nbslash\,T^A
          q(u\bar n)
          |\bar B\rangle}{2M_B} \nonumber\,.
\end{eqnarray}
Using the modeling of the soft function as in \cite{Benzke:2010js} allows us to 
estimate the size of resolved photon contributions. We need to estimate each of the $\tilde\Lambda_{ij}$ in (\ref {resolved}).

In order to estimate  $\tilde\Lambda_{78}^{\bar B}$, one can use Fierz transformation and the Vacuum Insertion Approximation (VIA) to express $h_{78}^{(1)}$ as a the square  of B meson light-cone amplitudes $\phi_+^B$. This allows us to write
$$ 
\tilde\Lambda_{78}^{\bar B}\bigg|_{\rm VIA}=e_{\rm spec}\,\frac{2f_B^2\,M_B}{9}\int_{0}^\infty\,
d\omega_1\,\frac{\left[\phi_+^B(\omega_1,\mu)\right]^2}{\omega_1}\,,
$$
where $e_{\rm spec}$ denotes the electric charge of the spectator quark in units of $e$ ($e_{\rm spec}=2/3$ for $B^-$ and $-1/3$ for $\bar B^0$). Using  \cite{Lee:2005gza} to constrain the integral, one finds that in the VIA, $\tilde\Lambda_{78}^{\bar B}\in e_{\rm spec}[17\,\mbox{MeV},\, 190\,\mbox{MeV} ]$.

Both $\tilde\Lambda_{17}^u$ and  $\tilde\Lambda_{17}^c$ depend on $h_{17}$. Being a soft function, $h_{17}$ has support over a hadronic range.  Since in the expression for $\tilde\Lambda_{17}^c$, the integral starts at  at $4m_c^2/m_b\approx 1$ GeV, we can expect a small overlap.  Indeed one finds that \cite{inprep}, $- 9\,\mbox{MeV} < \tilde\Lambda_{17}^c < + 11\,\mbox{MeV}$. For $\tilde\Lambda_{17}^u$ there is no such suppression and we have  $- 330\,\mbox{MeV} < \tilde\Lambda_{17}^u < + 525\,\mbox{MeV}
$. The range is not symmetric since the normalization of $h_{17}$  is $2\lambda_2\approx 0.24\, {\rm GeV}^2$. This is the same result as one would obtain from estimating $ \tilde\Lambda_{17}^u$ using naive dimensional analysis.

Including both the direct and resolved contributions and using  $\mu=2$\,GeV for the factorization scale,  we find that the total CP asymmetry in the SM is
$$
   {\cal A}_{X_s\gamma}^{\rm SM}
   \approx \pi\,\bigg|\frac{C_1}{C_{7\gamma}}\bigg|\,\mbox{Im}\,\epsilon_s\,
    \bigg( \frac{\tilde\Lambda_{17}^u-\tilde\Lambda_{17}^c}{m_b}
    + \frac{40\alpha_s}{9\pi}\,\frac{\Lambda_c}{m_b} \bigg) 
   = \left( 1.15\times\frac{\tilde\Lambda_{17}^u-\tilde\Lambda_{17}^c}{300\,\mbox{MeV}}
    + 0.71 \right) \% \,.
$$
The direct contribution is slightly higher then the $0.5\%$ mention before, since we are using a slightly lower factorization scale. The conclusion is that the asymmetry is actually dominated by non-perturbative effects.  Using the above estimates for $\tilde\Lambda_{17}^u$ and  $\tilde\Lambda_{17}^c$, we find that the CP asymmetry in the SM can be in the range
$-0.6\%<{\cal A}_{X_s\gamma}^{\rm SM}<2.8\%$. 

Beyond the SM, where the Wilson coefficients can be complex, we find that the asymmetry should be 
\begin{eqnarray}
   \frac{{\cal A}_{X_s\gamma}}{\pi}
   &\approx& \left[ \left( \frac{40}{81} - \frac{40}{9}\,\frac{\Lambda_c}{m_b} \right) 
    \frac{\alpha_s}{\pi} 
    + \frac{\tilde\Lambda_{17}^c}{m_b} \right] \mbox{Im}\,\frac{C_1}{C_{7\gamma}}- \left( \frac{4\alpha_s}{9\pi} - 4\pi\alpha_s\,e_{\rm spec}\,
    \frac{\tilde\Lambda_{78}}{m_b} \right) \mbox{Im}\,\frac{C_{8g}}{C_{7\gamma}} \nonumber\\
   &&- \left( \frac{\tilde\Lambda_{17}^u - \tilde\Lambda_{17}^c}{m_b}
    + \frac{40}{9}\,\frac{\Lambda_c}{m_b}\,\frac{\alpha_s}{\pi} \right)
    \mbox{Im}\bigg(\epsilon_s\,\frac{C_1}{C_{7\gamma}}\bigg)\,. 
\end{eqnarray}
Notice that the second term in this equation depends on the flavor of the spectator quark. In other words, the CP asymmetry can be different for charged and neutral B's. This effect arises already at order $\Lambda_{\rm QCD}/m_b$ for the resolved photon contribution. For the direct photon contribution such effects are more power suppressed. This allows us to suggest a new test of physics beyond the SM   by measuring the CP asymmetry difference
\begin{equation}
 {\cal A}_{X_s^-\gamma} - {\cal A}_{X_s^0\gamma}
    \approx 4\pi^2\alpha_s\,\frac{\tilde\Lambda_{78}}{m_b}\,
    \mbox{Im}\,\frac{C_{8g}}{C_{7\gamma}} 
   \approx 12\% \times \frac{\tilde\Lambda_{78}}{100\,\mbox{MeV}}\,
     \mbox{Im}\,\frac{C_{8g}}{C_{7\gamma}}\,. 
\end{equation}

We conclude with several comments about $\bar B\to X_{d}\,\gamma$.  All the above expressions apply also to this decay mode. All that we need to do is replace  $\epsilon_s$ by $\epsilon_d=(V_{ub} V_{ud}^*)/(V_{tb} V_{td}^*)=(\bar\rho-i\bar\eta)/(1-\bar\rho+i\bar\eta)$. As a result the CP asymmetry is enhanced by a factor of $\mbox{Im}(\epsilon_d)/\mbox{Im}(\epsilon_s)\approx -22$.  Including resolved photon contributions we find an asymmetry in the range $-62\%<{\cal A}_{X_d\gamma}^{\rm SM}<14\%$. Another quantity of interest is the untagged CP asymmetry for $\bar B\to X_{s+d}\,\gamma$.  Up to tiny U-spin breaking corrections, the direct photon contribution to the untagged asymmetry vanishes in SM \cite{Soares:1991te,Kagan:1998bh,Hurth:2001yb}. This result does not change even after including 
 resolved photon effects.

\section*{Acknowledgments}
I would like to thank organizers for the invitation to give a talk at Moriond QCD 2011. This work is supported by DOE grant DE-FG02-90ER40560.

\end{document}